\documentclass[twocolumn,amsmath,amssymb,floatfix,showpacs]{revtex4}

\usepackage{graphicx}
\usepackage{bm}
\usepackage{subfigure}

\newcommand{\be}{\begin{equation}}
\newcommand{\ee}{\end{equation}}
\newcommand{\M}{\mathcal{M}}
\newcommand{\mink}{\mathbb{M}^4}
\newcommand{\Lob}{\mathbb{H}^3}
\newcommand{\wE}{\widetilde{E}}
\newcommand{\wrho}{\widetilde{\rho}}

\begin{document}

\title{Energy-momentum diffusion from spacetime discreteness}

\author{Fay Dowker}
\author{Lydia Philpott}
        \email{l.philpott06@imperial.ac.uk}
\affiliation{Blackett
Laboratory, Imperial College, London SW7 2AZ, UK}
\author{Rafael D. Sorkin}
\affiliation{Perimeter Institute for Theoretical Physics,
Waterloo ON, Canada}

\begin{abstract}
We study potentially observable consequences of spatiotemporal
discreteness for the motion of massive and massless particles.
First we describe some simple models for the motion of a
massive point particle in a fixed causal set background. If the
causal set is faithfully embeddable in Minkoswki spacetime, the
models give rise to particle motion in the continuum spacetime.  
At large scales, the microscopic swerves induced by the underlying
atomicity manifest themselves as a Lorentz invariant diffusion in
energy-momentum governed by a single phenomenological parameter,
and we derive in full the corresponding diffusion equation.
Inspired by the simplicity of the result, we then derive the most
general Lorentz invariant diffusion equation for a massless particle,
which turns out to contain two phenomenological parameters describing,
respectively, diffusion and drift in the particle's energy.
The particles do not leave the light cone however: their worldlines
continue to be null geodesics.  Finally, we deduce bounds on the drift and
diffusion constants for photons from the blackbody nature of
the spectrum of the cosmic microwave background radiation.
\end{abstract}

\maketitle

\section{Introduction}

The search for a theory of quantum gravity is not, as yet, motivated by
experimental results. We currently have no
unambiguously relevant quantum gravitational phenomena to guide us
in developing candidate theories,
though it has long been suggested that
a nonzero value of the cosmological constant of order $10^{-120}$
could have a quantum gravitational
origin \cite{Sorkin:1990bh,Sorkin:1990bj,Sorkin:1997gi,Ahmed:2002mj}.
Outside of cosmology,
black hole thermodynamics is often mentioned as
one example of a realm where concepts of general relativity and
quantum mechanics must both come into play - but experimental black hole
physics is out of our reach for now and even
if analogue models of black holes in condensed matter
systems could be tested this would only probe the
semiclassical regime and not full quantum gravity.
The existing approaches to quantum gravity have therefore been developed
with the hope that the confrontation with experiment can be postponed.
At the present time, however,
the growing number of different approaches
means that the importance of testing ideas against
observation, if at all possible, is greater than ever.

Experimental verification
of  quantitative and unexpected predictions is of the utmost
importance in the development of a successful new theory.
An example pertinent to the current paper is
 if we were to find observational evidence that spacetime
is fundamentally discrete, then that would have a major
impact on the direction of quantum gravity research.
What form might such
evidence take; what could be the Brownian motion of
our age? To answer that question requires the development of
phenomenology that draws on essential aspects
of a discrete theory of quantum gravity
which turns out to be achievable in the causal set approach.

Causal set theory is a discrete, Lorentz invariant approach to
quantum gravity \cite{Bombelli:1987aa,tHooft:1979, Myrheim:1978}.
For reviews and further references see, for example,
\cite{Henson:2006kf,Dowker:2006sb,Sorkin:2003bx}.
It is a work in progress: a quantum causal set
dynamics still eludes us. Without a quantum dynamics it
seems at first sight premature to develop causal set
phenomenology but the kinematics of
causal set theory is so concrete that we are able to make
some progress in this direction.

A causal set is
a locally finite partial order and is the kinematical basis
for the theory. One could state the central hypothesis as that
spacetime {\textit {is}} a causal set, or, if one wanted to
hedge one's bets whilst the foundations of quantum theory are
laid, that causal sets are the histories in a sum-over-histories
quantum theory of spacetime.

In detail, a causal set is a set $C$ endowed with a binary
relation $\prec$ satisfying:
\begin{enumerate}
        \item transitivity: if $x\prec y$ and $y\prec z$ then $x\prec
        z$, $\forall x,y,z\in C$;
        \item reflexivity: $x\prec x$, $\forall x \in C$;
        \item acyclicity: if $x\prec y$ and $y \prec x$, then $x=y$,
$\forall x, y \in C$;
        \item local finiteness: $\forall x, z\in C$ the set
        $\left\{y\mid x\prec y\prec z\right\}$ of elements is finite.
\end{enumerate}

Our
observed continuum Lorentzian manifold, it is assumed,
arises as an approximation to an underlying causal set.
The partial order gives rise to the causal ordering of events in the
approximating continuum spacetime,
and the number of elements comprising a spacetime region gives the volume
of that region in fundamental units which we take to be of order the Planck
volume.
The above rules of correspondence give an essentially unique way to
associate a class of causal sets to a given continuum spacetime via
a process known as
``sprinkling'' defined as follows.
Given a Lorentzian manifold, $(M,g)$,
points are selected from $M$ randomly
via a Poisson process in which the probability measure is
equal to the spacetime volume measure in some fundamental units.
The selected points are the elements of a causal set
once they have been endowed with the
partial order induced by the spacetime causal order.
The number of
points chosen from any region of the manifold
will be approximately equal to
the volume of the region (in
fundamental units) up to Poisson fluctuations. For more
details on sprinklings see the reviews mentioned above,
for a proof of the Lorentz invariance of the
process see
\cite{Bombelli:2006nm}.
We thus have a straightforward way to construct a causal set that could be
the discrete underpinning of a particular continuum spacetime
and, consequently, a starting point to develop the
phenomenology of discrete spacetime.

An obvious place to look for consequences of causal set theory is in
the behaviour of particles. If the underlying spacetime is a discrete
structure rather than a continuous manifold, free particles
might no longer be able to follow precise timelike geodesics.
Intuitively, the underlying discreteness could cause the
particles to `swerve'
and indeed a model of particle behaviour illustrating this
was proposed in~\cite{Dowker:2003hb}.
There, a classical particle is modelled in the simplest possible way,
as a point with no internal structure.
The causal set, $C$, considered is a sprinkling into
Minkowski spacetime and a particle trajectory
consists of a \textit {chain} of elements where a
chain is a totally ordered subset of $C$.
The trajectory is constructed iteratively, where the
trajectory's past determines its future, but only a certain proper
time $\tau_f$ (the ``forgetting time'') into the past is relevant. If
the particle has reached an element $e_n$ with four-momentum $p_n$,
the next element $e_{n+1}$ is chosen such that
\begin{itemize}
\item $e_{n+1}$ is in the causal future of $e_n$ and within a proper
time $\tau_f$,
\item the momentum change $\left|p_{n+1} - p_n\right|$ is minimised.
\end{itemize}
Here the momentum $p_{n+1}$ is defined to be proportional to the
vector between $e_n$ and $e_{n+1}$ and on the mass shell. Heuristically,
the trajectory tries to stay as straight as possible at each
step. 
 Indeed, an alternative way to define the process is to 
 specify that in a frame in which the last two elements of
 the trajectory, 
 $e_{n-1}$ and $e_n$, lie on the vertical $t$-axis, the next
 element $e_{n+1}$ is chosen to be the one, within 
 proper time $\tau_f$ to the future of $e_n$, such that 
 the vector from $e_n$ to $e_{n+1}$ is as close to vertical as 
 possible. 

 In this simple model the discreteness of a causal set
 results in random fluctuations in the momentum of a particle.

One could object that this model is not intrinsic to the causal set as it
makes use of information in the continuum manifold to define the
momentum change. However, similar models can be defined with no
reference to the continuum. Two such models are proposed
in Section~\ref{s:intrinsicmodels}.
One of our main claims is that, whatever the microscopic model
of particle motion, if it is Lorentz invariant and
gives rise to small random fluctuations in the momentum
of the particle then it can be approximated by a
continuum description as a diffusion in momentum space.
In Section~\ref{s:massiveparticles} we
support this claim by
giving the derivation of the diffusion equation for massive
particles introduced in \cite{Dowker:2003hb}. We
also derive the
particle diffusion equation in a more useful cosmic time form
and without the original assumption of spatial homogeneity.

In Section~\ref{s:masslessparticles} we explore the case of massless
particles on a causal set and obtain diffusion equations for the
momentum of massless particles in the continuum approximation.
Bounds are placed on
the constants in the massless particle diffusion equation in
Section~\ref{s:bounds} by considering the effect of momentum diffusion
on the spectrum of the cosmic microwave background.  We will use
units in which $c= h = G = 1$ --- which we will refer to as ``Planck
units''. Fundamental units are
related to Planck units by a, yet to be determined factor of order 1.
Boltzmann's constant
is also set to one, $k_B = 1$.

\section{Intrinsic models for massive particles}
\label{s:intrinsicmodels}

As mentioned above, the original microscopic model in
\cite{Dowker:2003hb} depended on information from the
continuum Minkowski spacetime whereas a better model ought
to be intrinsic to the causal set itself and rely only on the
order relation.
Two slightly different intrinsic models will be described in this
section, to give an idea of the wealth of possibilities available.

We first recall some causal set definitions.
Let $C$ be a causal set.
\begin{itemize}
\item A \emph{link} is an irreducible relation, i.e.~a pair of
distinct
elements
$a,b$ such that $a \prec b$ and
there exists no distinct $c$ such that $a\prec c \prec b$.
\item A \emph{chain} is a totally ordered subset of $C$. An \emph{n chain}
is a chain with $n$ elements and its \emph{length} is $n-1$, the number
of links.
\item A \emph{longest chain} between two
elements is a chain
whose length is maximal amongst chains between those endpoints.
There may be more than one
longest chain between two elements.
\item On a causal set the closest approximation we have to a timelike
geodesic between two elements is a longest chain.
For two causal set elements $a$ and $b$ the
length of a longest chain between $a$ and $b$ will be denoted
$d(a,b)$. For sprinklings into Minkowski spacetime, in the
asymptotic limit of large distances, $d(a,b) \sim \alpha T$ where $T$ is
the proper time between $a$ and $b$ and $\alpha$ is a (dimension
dependent) constant \cite{Brightwell:1991}.
\item There is a link between elements $a$ and $b$ iff
$d(a,b) = 1$.
\item
A \emph{path} is a chain consisting entirely of links, i.e.~a set of
elements $a\prec b\prec c\prec d\prec\ldots$ such that $d(a,b) = 1$,
$d(b,c) = 1$, $d(c,d) = 1\ldots$.
\end{itemize}

\begin{figure*}[th]
\begin{center}
\subfigure[]{\label{f:model1}
\includegraphics[height = 4cm]{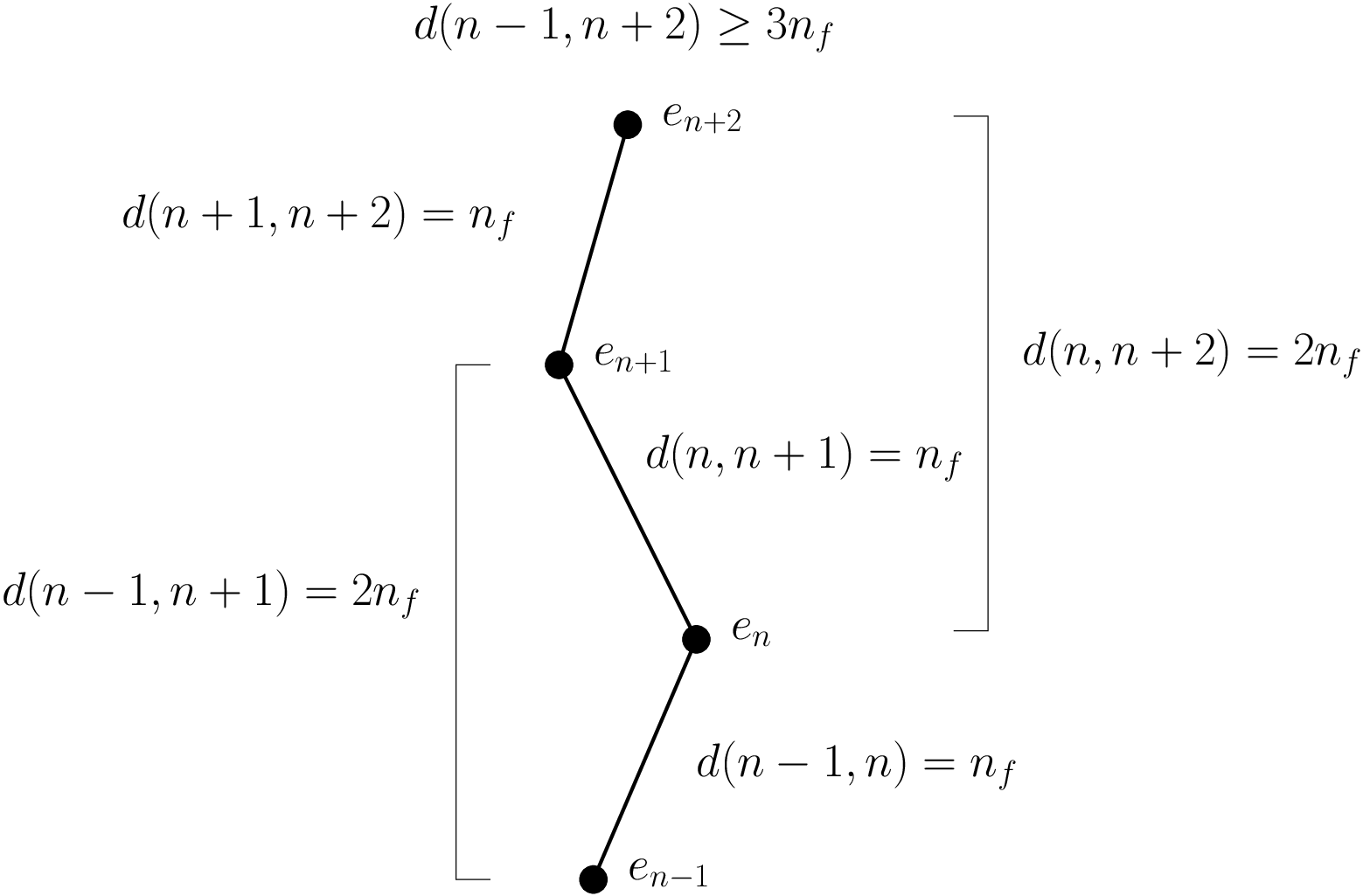}}
\hspace{20mm}
\subfigure[]{\label{f:model2}
\includegraphics[height = 4cm]{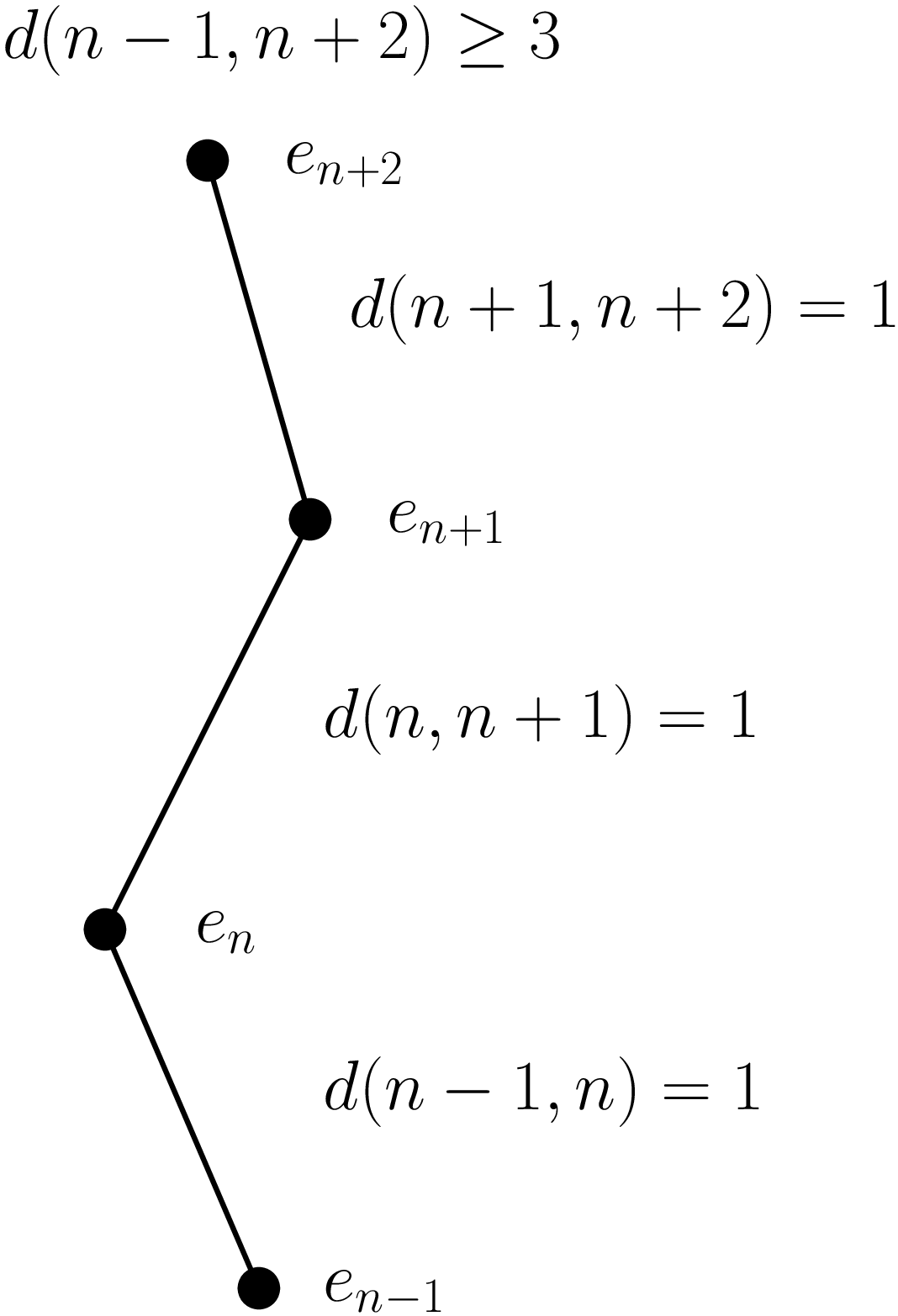}}
\caption{A trajectory constructed using (a) model 1 and (b) model 2.}
\end{center}
\end{figure*}

\subsection{Model 1}
If
a dynamical rule for particle motion is to be intrinsic to
the causal set background it can no longer refer to a forgetting time
$\tau_f$. This instead becomes a `forgetting number', an integer
$n_f >> 1$. In Intrinsic Model 1 a particle trajectory is
a chain, $\dots e_{n-2} \prec e_{n-1} \prec e_n \dots$ which is
determined by the following (Markov of order 2) process:

 Given a partial particle trajectory $\ldots e_{n-1}, e_n$ the next
element $e_{n+1}$ is chosen such that
\begin{itemize}
\item $d(e_n,e_{n+1})=n_f$,
\item $d(e_{n-1},e_{n+1}) = 2n_f$,
\end{itemize}
(see Figure~\ref{f:model1}).  These requirements do not guarantee the
existence of a unique such $e_{n+1}$. However there will almost surely
be finitely many eligible elements and
 we therefore construct the
trajectory by choosing an element uniformly at random from these.

The particle trajectory should swerve a little, but remain
approximately straight so long as $n_f$ is large, since,
in that case, the results of Brightwell and Gregory \cite{Brightwell:1991}
show that the expected position of $e_{n+1}$ is close to the hyperboloid
of points proper distance $n_f/\alpha$ from $e_n$ and
to the  hyperboloid
of points proper distance $ 2 n_f/\alpha$ from $e_{n-1}$.

In this model we can consider the trajectory
as consisting of just the elements $\ldots e_{n-1}, e_n, e_{n+1}\ldots$
or of the ``filled in chain'' consisting of  a (randomly chosen)
longest chain (of length $n_f$) between $e_{n-1}$ and
$e_n$, another longest chain between $e_n$ and $e_{n+1}$ (also length
$n_f$) and so on. By imposing $d(e_{n-1},e_{n+1}) = 2n_f$ we have
forced the chain of length $2n_f$ that we have between $e_{n-1}$ and
$e_{n+1}$ also to be a longest chain. The trajectory is thus
approximately geodesic over all $\left\{e_{n-1}:e_{n+1}\right\}$
segments. The trajectory consisting of longest chains between
$e_{n-1}$ and $e_n$, $e_n$ and $e_{n+1}$, and $e_{n+1}$ and $e_{n+2}$
is not, however, necessarily a longest chain between $e_{n-1}$ and
$e_{n+2}$.

Possible variations on this model include
choosing, at random, the forgetting number at each step so that
the mean is $n_f$ with some fixed variance.

\subsection{Model 2}

The trajectory is explicitly constructed as a path in this model,
i.e. $d(e_n,e_{n+1})=1$ for all $n$. Given a partial particle
trajectory $\ldots e_{n-n_f}, \ldots, e_{n-1}$ the next element
$e_{n}$ is chosen such that
\begin{itemize}
\item $d(e_{n-1},e_{n})=1$,
\item $d(e_{n-n_f},e_{n}) + \ldots + d(e_{n-2},e_{n}) + d(e_{n-1},
e_{n})$ is minimised,
\end{itemize}
(see Figure~\ref{f:model2}).  Note that this minimisation does not
necessarily yield a unique $e_{n}$, in which case we construct the
trajectory by choosing an element uniformly at random from those eligible. Also,
if the past trajectory has length less than $n_f$ the minimisation is done
over all elements available.

Each element is linked to the previous, i.e. $d(e_{n-1}, e_{n}) =
1$ so we know there exists a chain (our trajectory) of length $n_f$
between $e_{n-n_f}$ and $e_n$. The maximal chain length,
$d(e_{n-n_f},e_n)$, must therefore be greater than or equal to $n_f$. If we
choose $e_n$ to minimise $d(e_{n-n_f},e_n)$ we ask that the trajectory
be as close as possible to geodesic between $e_{n-n_f}$ and $e_n$
while fulfilling $d(e_{n-1}, e_{n}) = 1$. Minimising the sum of
the partial lengths distributes the geodesic property along the
path.

\section{The continuum approximation for massive particles}
\label{s:massiveparticles}
The models  described above 
are intrinsic to a causal set and as such could be used to 
define particle motion on any causal set whatsoever. For 
phenomenology however the models are only of interest when  
defined on a causal set that arises by sprinkling into 
four-dimensional Minkowski spacetime, $\mathbb{M}^4$, or some
other physical spacetime such as a Friedmann-Robertson-Walker (FRW) cosmology. 
It is the central conjecture, or Hauptvermutung, 
of causal set theory that  
a causal set that arises by sprinkling into $\mathbb{M}^4$,  
is well-approximated by $\mathbb{M}^4$ and that the 
associated embedding of the causal set in $\mathbb{M}^4$  gives us a way to
derive spacetime physics from physics on the causal set.
The Hauptvermutung remains to be
proved -- though we have a quite a bit of
evidence for it especially in the case of flat spacetime (see, for example~\cite{Brightwell:1991,meyer:1988})
--  and in the context of the present work 
it is the central phenomenological assumption.

With this assumption,
the models of Section~\ref{s:intrinsicmodels} defined on 
a causal set which arises from sprinkling
into $\mathbb{M}^4$ are also models of particle motion in 
$\mathbb{M}^4$. To be completely
explicit: in both models the particle's motion on the causal set --
a chain --
defines a piecewise linear trajectory in $\mathbb{M}^4$ in the following 
way. By assumption, the causal set arises from 
sprinkling and so is embedded in $\mathbb{M}^4$ (in such a way that the 
order on the elements respects the causal order on the embedded
positions and the number of elements in any region 
is approximately the spacetime volume in fundamental units). Between
two causal set elements in the chain of the
particle's motion on the causal set, its \textit{spacetime} trajectory 
through $\mathbb{M}^4$ is
a straight line and its four-momentum is well-defined by requiring
it to be on the mass shell, hyperbolic space $\mathbb{H}_3$.

The original swerves model can be similarly converted into a model of 
piecewise linear particle worldlines in $\mathbb{M}^4$ and we can consider 
each of these spacetime models as defining an evolution on a manifold
of states. The manifold of states
for the particle is $\mathbb{M}^4\times \mathbb{H}_3$ -- its position in space
and its momentum -- and the time parameter for the evolution is 
proper time. This evolution is effectively
stochastic because although knowledge of the causal set makes the
trajectory deterministic, we treat the causal set as unknown,
the analogue of the state of the water molecules causing Brownian motion.

If we define $n_{\textrm{macro}}$ to be the scale of macroscopic
physics measured in Planck units then 
we can demand a separation of scales
so that $1<< n_f << n_{\textrm{macro}}$. Indeed, the Planck scale is so 
small that even if we want to choose the scale of Large Hadron Collider 
physics as our ``macro'' scale ($n_{\textrm{macro}} = 10^{15}$)
there's still plenty of room to chose $n_f$.  
The
change in spacetime position which is 
bounded by $n_f$ Planck units will then be small  at each step  
as will the change in momentum.

The dynamics is therefore a stochastic evolution on a manifold
of states and such systems are dealt with in general 
in the formalism developed by Sorkin \cite{Sorkin:1986}.  
At the macroscopic scale of many ($n_{\textrm{macro}}/n_f$)
steps the process can be described approximately as a diffusion.

Although the models
described above cannot be considered completely
realistic (for example the particles are
classical and zero-size) we claim that provided
the process is Lorentz- and translation-invariant,
it will always give rise to the same diffusion
equation, namely the equation written down in \cite{Dowker:2003hb}.
As promised there,
we present below the full derivation of this equation,
supporting our claim that the continuum model is universal and
independent of the discrete microscopic details.
We derive the equation initially with the particle's proper time playing
the role of independent variable; we then obtain the equivalent
equation in terms of cosmic time, by expressing both in terms of a
conserved current in a certain space of eight dimensions.
This Lorentz invariant process was first considered
by Dudley \cite{Dudley:1965,Dudley:1967}, though one of his
diffusion equations conflicts with ours.
Without any imposition of Lorentz invariance,
a general formalism for describing
diffusion in Minkowski space was set up
by Schay \cite{Schay:1961}.
%

\subsection{The diffusion equation for a massive particle}

We use the general formalism of \cite{Sorkin:1986}, which deals with
stochastic evolution on a manifold of states.
The state space, $\mathcal{M}$, of the swerving particle of mass $m$ is
$\mathcal{M} = \mathbb{M}^4\times\mathbb{H}^3$, where $\mathbb{H}^3$ is
the mass shell.
The coordinates on $\mathbb{M}^4$ are the usual
Cartesians $\{x^\mu \}$, $\mu = 0,1,2,3$ and indices are
raised and lowered with $\eta_{\mu\nu}$, the Minkowski metric.
The spatial coordinates
on $\mathbb{M}^4$ will be written
as $\{x^i\}$. Cartesian coordinates in momentum
space are $p_\mu$ and whenever they are used it will be
understood that $p_\mu$ lies on the mass shell which is the
hyperboloid in momentum space defined by $p_\mu p^\mu + m^2 = 0$.
$p^0 = E$ is the energy and $p = \sqrt{p_1^2 + p_2^2 + p_3^2}$ is the
norm of the three momentum.
 The three coordinates on $\mathbb{H}^3$ will be written abstractly as
$p^a$.
We denote the coordinates on $\M$ collectively as
$X^A = \{ x^\mu, p^a\}$ and in what
follows capital letters $A, B$ will be used to
indicate general indices on $\mathcal{M}$; $\mu, \nu$ are indices on
$\mathbb{M}^4$; $i,j$ are spatial indices on $\mathbb{M}^4$;
$a, b$ are indices on $\mathbb{H}^3$.

The metric on $\M$ is the product of the Minkowski
metric $\eta_{\mu\nu}$ on $\mink$ and the Lobachevski metric
$g_{ab}$ on $\Lob$.
This is the unique Poincar\'e invariant metric (up to an overall constant).
The ``density of states'', $n$,
plays a role in the formalism of \cite{Sorkin:1986}, and by symmetry, it
must be proportional
to the volume measure on $\M$, so $n \propto \sqrt{g}$ where
$g = {\textrm{det}}(g_{ab})$. The ``entropy scalar'', $s$, is
given by $s = \ln (n)$
(Boltzmann's constant has been set to 1).\footnote%
{For present purposes the identification of $n$ with a density of
 microscopic states is unnecessary.  What is relevant is that a
 probability-density proportional to $n$ be in equilibrium (i.e.
 time-independent).}

A process that undergoes stochastic evolution on a manifold of states,
$\mathcal{M}$, in time parameter $T$,
can be described by a current, $J^A$ and a continuity
equation~\cite{Sorkin:1986}:
\begin{eqnarray}
J^A &=& -\partial_B\left(K^{AB}\rho\right)+v^A\rho\label{e:current}\,,\\
\frac{\partial\rho}{\partial T} &=& -\partial_A
J^A .\label{e:continuity}
\end{eqnarray}
Here the probability density for the system is given by
$\rho=\rho\left(X^A,T\right)$, a scalar density on
$\mathcal{M}$.  The coefficients $K^{AB}$  are given by
\be\label{e:Kdef}
K^{AB} = \lim_{ \Delta T \to 0+} \left< \frac{\Delta X^A \Delta X^B}{2\Delta T} \right>\, ,
\ee
where $<\cdot>$ denotes expectation value in the process in which the particle
starts at a definite point of $\M$ (page 146 of \cite{Sorkin:1986}).
$K^{AB}$ is
a symmetric, positive semi-definite
matrix which transforms as
the components of
a tensor on $\M$.
The coefficients $v^A$ are
\be\label{e:vdef}
v^A = \lim_{ \Delta T \to 0+} \left< \frac{\Delta X^A}{\Delta T} \right>\, ,
\ee
and do not transform as a
vector on $\M$, but can be combined with $K$ and the entropy scalar $s$ to
form a true vector $u^A$,
\begin{equation}
u^A = v^A - \partial_B K^{AB} - K^{AB}\partial_B s\,.\label{e:uA}
\end{equation}

The current and
continuity equations can be reexpressed in terms of the true vector
$u^A$:
\begin{equation}
 \frac{\partial\rho}{\partial T} =
 \partial_A\left(K^{AB}n\partial_B\left(\frac{\rho}{n}\right)-u^A\rho\right)
 \label{e:diffequA}.
\end{equation}
To find the diffusion equation for our particle process,
therefore, we need to determine $K^{AB}$ and
$u^A$.

Requiring the equation to be Poincar\'e invariant is a
very stringent condition and proves to be sufficient
for us to determine $K^{AB}$ and $u^A$, up to
the choice of one constant parameter. This means that the
resulting equation is very robust and independent of the
details of the underlying particle model so long as it is
Poincar\'e invariant.

Consider the process
referred to
$\tau$,
proper time along the worldline
of the particle. Then
\be
K^{\mu\nu} = \lim_{ \Delta \tau \to 0+} \left< \frac{\Delta x^\mu \Delta x^\nu}{2\Delta \tau}
\right>\, .
\ee
$\Delta x^\mu = \frac{1}{m} p^\mu \Delta  \tau$ at every step of the process and
so $K^{\mu\nu} = \frac{1}{2}\lim_{\Delta \tau \to 0} p^\mu p^\nu \Delta \tau = 0$.
Since $K^{AB}$
is positive definite, this implies that $K^{\mu A} = 0$ and the only nonzero
components are $K^{ab}$. The only Lorentz invariant tensor on $\Lob$ is
proportional to the metric, $g^{ab}$ and the coefficient is independent of
$x^\mu$ by translation invariance. So we have
\begin{equation}
K^{AB}= \left(
\begin{array}{cc}
0 & 0\\ 0 & kg^{ab}
\end{array}
\right)\,,
\end{equation}
where $k>0$ is a constant.

Now consider
\be
v^\mu = \lim_{ \Delta \tau \to 0+} \left< \frac{\Delta x^\mu}{\Delta \tau} \right>\, ,
\ee
which, by the above is $v^\mu = p^\mu/m$.
The components of the true vector $u^\mu$ are
equal to $v^\mu$ because
$K^{\mu A} = 0$.
There is no Lorentz invariant vector on
$\Lob$ and so $u^a = 0$:
\be
u^A=\left(p^\mu/m,0\right)\,.
\ee
We can now write down  the proper time diffusion equation from
(\ref{e:current}) and (\ref{e:continuity}):
\be \label{e:mtau}
\frac{\partial\rho_{\tau}}{\partial\tau} =
k \; \partial_a \left( g^{ab} \sqrt{g} \partial_b \left(\frac{\rho_\tau}{\sqrt{g}}
\right)\right) - \frac{1}{m} p^\mu \partial_\mu \rho_\tau\;.
\ee

If we define a scalar $\overline{\rho} = \rho_\tau/\sqrt{g}$ we obtain
the equation in reference \cite{Dowker:2003hb}:
\be
\frac{\partial\overline{\rho}}{\partial\tau} = k\;\nabla^2_H \overline{\rho}
-  \frac{1}{m} p^\mu \partial_\mu \overline{\rho}\,,
\ee
where $\nabla^2_H$ is the Laplacian on $\Lob$.

\subsection{Diffusion in cosmic time for massive particles}

\begin{figure}[th]
\begin{center}
\includegraphics[width = 0.35\textwidth]{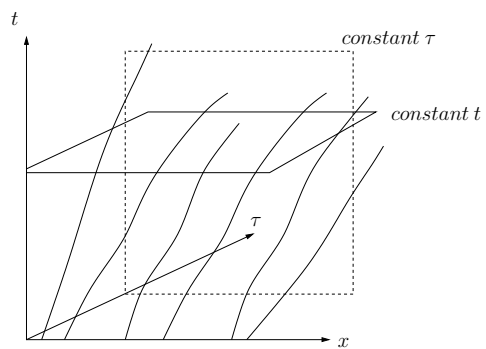}
\caption{Particle trajectories as flowlines in
$\M' = \mathbb{M}^4\times\mathbb{H}^3\times\mathbb{R}$ (where we have suppressed two spatial dimensions).}
\label{f:ttau}
\end{center}
\end{figure}

 Given an initial distribution of particles, for instance from an
 astronomical source, the above equation is not very useful for
predicting
 the results of observations. Even if particles all leave the
 source at the same
 time
with the same momentum, the momentum variation
 induced by the swerves will result in particles arriving after different
 proper times and at different observatory times. The proper time that
elapses along the particles' worldlines from source to
detector is not
observable.
 To compare the swerves
 model with experiment and observation it is necessary to describe the
 evolution of the distribution in time in the rest frame
of our detector, which time we refer to as {\it cosmic time}.

 A first step in this direction was to look at the
 nonrelativistic limit of the proper time diffusion equation,
 when
 proper time and cosmic time are comparable. The nonrelativistic limit
 in fact proves sufficient to place very strong bounds on the value of
 the diffusion constant and severely limit any observable effects (see~\cite{Dowker:2003hb, Kaloper:2006pj}).

In the fully relativistic case, Dowker et al.~wrote down the diffusion
 equation in terms of cosmic time for the special case of an initially
 spatially homogeneous distribution~\cite{Dowker:2003hb}.
We will now give the derivation of the cosmic time evolution equation for
the general case of spatially inhomogeneous distributions.

The conversion between proper time and cosmic time
is possible because both are good time parameters along
all possible particle worldlines, which are
causal. If we visualise our diffusion
process as a collection of such worldlines through
spacetime and momentum space,
both cosmic time, $t = x^0$, in our chosen frame and proper time $\tau$
increase monotonically along each trajectory.
Adding proper time to our state space by assuming that the particle
starts at parameter $\tau = 0$ and cosmic time $t=0$,
the process is represented by
flowlines in
$\M' = \mathbb{M}^4\times\mathbb{H}^3\times\mathbb{R}$ (see
Figure~\ref{f:ttau}) and along each flowline,
both $\tau$ and $t$ are good time parameters. The proper time diffusion
equation we have found describes the evolution of the distribution on
constant $\tau$
hypersurfaces in $\M'$.
What we want is to obtain the diffusion equation
for evolution of the distribution on constant $t$ hypersurfaces
integrated over all proper times.

First we put $t$ and $\tau$ on an equal footing by considering
the larger space $\M'$ and defining
a new current component
\begin{equation}
J^{\tau}(t,x^i,p^a,\tau) = \rho_{\tau}\;.
\end{equation}
If we denote coordinates on this extended space,
$\M'= \M \times\mathbb{R}$, by $X^\alpha = \{X^A, \tau\}$
then the continuity equation (\ref{e:continuity}) can be written
\be
\partial_\alpha J^\alpha=0\;.
\ee

Using equation (\ref{e:current})
(and still treating $\tau$ as our time-parameter)
 we can express the $t$ component of the current
in terms of $J^{\tau}$ (a.k.a. $\rho_{\tau}$).
\begin{eqnarray}
J^{t}(t,x^i,p^a,\tau)
&=& -\partial_B\left(K^{tB}J^{\tau}\right)+v^tJ^{\tau}\nonumber\\
&=& v^t J^{\tau}\nonumber\\
&=& \gamma J^{\tau}\,,
\end{eqnarray}
where $\gamma = \partial t/ \partial \tau$
is the usual relativistic gamma factor.
The remaining components of the current can now be written in terms of
$J^t$. The spatial components are:
\begin{eqnarray}
J^{i}(t,x^i,p^a,\tau)
&=&-\partial_B\left(K^{iB}J^{\tau}\right)+v^iJ^{\tau}\nonumber\\
&=& v^i J^{\tau}\nonumber\\
&=& \frac{p^i}{m} \frac{J^t}{\gamma}.
\end{eqnarray}
In the case of the $p$ components the algebra is simpler if we first
note that we can express (\ref{e:current}) in the form
(cf. (\ref{e:diffequA}))
\begin{equation}
J^A = -K^{AB}\left(n\,\partial_B\left(\frac{\rho}{n}\right)\right) +
\rho u^A,
\end{equation}
 and so
\begin{eqnarray}
J^{a}(t,x^i,p^a,\tau)
&=& -kg^{ab}\, n \,\partial_b\left(\frac{J^t}{\gamma\,n}\right)\nonumber\\
&=& -kg^{ab}\sqrt{g}\partial_b\left(\frac{J^t}{\gamma
\sqrt{g}}\right)\;.
\end{eqnarray}
The metric $g^{ab}$ that appears here is
the Lobachevski metric on $\mathbb{H}^3$.

Since $\tau$ is unobservable
we need to integrate $J$
over $\tau$ and we denote the integrated current by $\bar{J}$.
Integrating the $t$ component of the current over proper time
from zero to infinity, gives
us the probability density on a hypersurface of constant t:
\begin{eqnarray}
\rho_{t}&=&\bar{J}^t(x^i,p^a,t)\nonumber\\
 &\equiv&\int{J^t d\tau}.
\end{eqnarray}
The components of the new current can be written:
\begin{eqnarray}
\bar{J}^{i}(x^i,p^a,t)&\equiv& \int{J^{i} d\tau}\nonumber\\
&=& \int{\frac{p^i J^t}{m \gamma} d\tau}\nonumber\\
&=& \frac{p^i}{m} \frac{\bar{J}^t}{\gamma}= \frac{p^i}{m} \frac{\rho_t}{\gamma}\,,\\
\bar{J}^{a}(x^i,p^a,t)&\equiv&\int{J^{a} d\tau}\nonumber\\
&=&-kg^{ab}n\partial_b\left(\frac{\bar{J}^t}{\gamma n}\right)\nonumber\\
&=& -kg^{ab}n\partial_b\left(\frac{\rho_t}{\gamma n}\right)
\;.
\end{eqnarray}

If we integrate the continuity equation over $\tau$ we obtain
\be
\left[J^\tau\right]_0^\infty + \partial_t \bar{J}^t + \partial_i \bar{J}^i +
\partial_a \bar{J}^a = 0\;.
\ee
$J^\tau|_{\tau =0}$ is zero for all $t>0$ and $J^\tau$ tends to zero as
$\tau$ goes to infinity for finite $t$. So for all $t>0$ we have
\be
\partial_t \bar{J}^t + \partial_i \bar{J}^i +
\partial_a \bar{J}^a = 0\,,
\ee
which gives the
 cosmic-time diffusion
equation
\begin{equation}\label{e:swervescosmic}
  \frac{\partial\rho_{t}}{\partial t} =
   -\frac{p^i}{m \gamma}\partial_i\rho_{t} +
   k\; \partial_a
   \left(g^{ab}\sqrt{g}\partial_b\left(\frac{\rho_{t}}{\gamma \sqrt{g}}\right)\right).
\end{equation}

This is a powerful phenomenological model because it depends on only
one parameter, the diffusion constant $k$. Data can therefore strongly
constrain $k$.

We note that this solves a problem posed by Dudley \cite{Dudley:1965}.
We also point out that Dudley's equation for the spatially
homogeneous distribution on page 267 of \cite{Dudley:1965}
is inconsistent with our equation (\ref{e:swervescosmic}).
(Equation (3.60) of \cite{Schay:1961} also
differs from the 1+1 dimensional analog of (\ref{e:swervescosmic}).)


\section{Massless particles}
\label{s:masslessparticles}
If
an
 underlying spacetime discreteness results in diffusion in momentum
and spacetime for massive particles, it is interesting to consider
whether a similar diffusion occurs for massless particles.
For massive
particles the concrete models for particle dynamics
on a causal set described above
motivated
the derivation of the diffusion equation (\ref{e:mtau}).
The case of massless particles on a causal set background
is rather different. If we consider a sprinkling into
Minkowski spacetime, for any given element, $p$, there
will almost surely be no element sprinkled on
 the future light cone of $p$.
The analogue of the future light cone of $p$ in
a sprinkling of Minkowski spacetime is the set of
all elements preceded by and linked to $p$. The elements are
distributed, roughly, near the hyperboloid one Planck unit of
proper time to the future of $p$.
Although the whole light cone
thus
has a good causal set analogue\footnote%
{Strictly speaking, it would be more correct to identify the ``light
 cone'' with a ``virtual boundary'' separating the future of $p$
 from the set of elements spacelike to it.},
the easiest analogue of
a null ray is only a single link,
making it hard to see how to
construct a discrete Markovian process
that
would result
in a close-to-null trajectory.

Modeling
the propagation of massless point particles on a causal set
as an approximately local process is
therefore problematic.  It is hoped that in the
future,
the study of massless
{\it fields}
on a causal set will enable us to model
massless particle propagation as wave packets, say.
In the meantime, however,
lack of knowledge of the exact nature of massless particle propagation on
the discrete level, does not mean we cannot derive a diffusion equation to
describe the potential effect of discreteness on photons
in the continuum approximation. We can arrive at a
massless diffusion equation in two ways: using the stochastic
evolution on a manifold of states procedure as for the massive
particle case, or simply taking a $m\rightarrow 0$ limit
of the diffusion equation for massive particles.
It turns out that the second method gives an
incomplete result.

The state space in the massless case differs from the massive
case.
For massive particles we had a probability distribution on
$\mathbb{M}^4\times\mathbb{H}^3$.
For massless particles $\mathbb{H}^3$ becomes the light cone in
momentum space defined by $p_\mu p^\mu = 0$.
This cone will be denoted $\mathbb{H}_0^3$.
If we assume that the photons under consideration are well described in
a geometrical optics approximation so they have definite spacetime
worldlines and momenta, our state space will be~\footnote
{More correctly this is the state-space for a massless particle of spin
 zero.  For a true photon, the state-space would be enlarged so as to
 describe also the polarization.}
$\mathbb{M}^4\times\mathbb{H}_0^3$.
Since proper time vanishes along a lightlike worldline, it is no longer
a suitable time parameter for our diffusion process.  We define,
instead, an affine time, $\lambda$, along any photon worldline by
$$
   dx^\mu = p^\mu d\lambda  \ .
$$
Notice that the normalization of this affine parameter is not
arbitrary.  It is fixed by its relation to the particle's four-momentum,
or geometrically, to its de Broglie wavelength.  Under the latter
interpretation, the affine parameter along a photon worldline $\gamma$
measures the area swept out in spacetime by a vector connecting $\gamma$
to a neighboring null geodesic that trails it by one wavelength.

In the massive particle case we equated the density of microstates, $n$,
to the determinant of the metric on our state space: $n\propto\sqrt{g}$.
In the massless case, the metric induced on $\Lob_0$
degenerates, but $\Lob_0$ still possesses a Lorentz invariant
measure of volume (unique up to a constant factor).
The four dimensional volume element $d^4 p$ of momentum space,
together with the masslessness constraint, $p^\mu p_\mu=0$,
lets us construct on $\Lob_0$ the invariant volume element
$d^4p\delta(p^\mu p_\mu)=d^3p/2p^0$,
i.e. $n\propto 1/p^0$ in Cartesian
coordinates.
It will be more useful, however, to work in polar coordinates
on $\Lob_0$: $\{p, \theta, \phi\}$
where $p$ is the magnitude of the three momentum
and $\theta$ and $\phi$ are the usual polar angles in momentum space.
In these coordinates, the density of states is
$n \propto p \sin\theta$.
There is also a
(unique up to a constant factor)
invariant
vector field
on
$\Lob_0$ which is the momentum itself, $p^a$, i.e.~the vector with components $(p, 0, 0)$ in polar coordinates.
This is absent in
the massive case,
where
 the momentum
vector does not lie in the mass shell.
Finally,
although there is no invariant metric on $\Lob_0$, there is
an invariant symmetric 2-tensor, $p^a p^b$
(unique up to a constant factor).

We first consider the process in affine time, $\lambda$.
As with the massive case,  we begin with the current and continuity equations,
(\ref{e:current}) and (\ref{e:continuity}), and determine $K^{AB}$ and
$u^A$.
Using the formulae (\ref{e:Kdef}) and (\ref{e:vdef}) with $T = \lambda$ we
find
\be
  K^{\mu\nu} = \lim_{\Delta \lambda \to 0+}  p^\mu p^\nu \Delta \lambda = 0.
\ee
$K$ is positive semidefinite so $K^{\mu a} = 0$,
 and finally
$K^{ab}$ must be Lorentz invariant and translation invariant so
\begin{equation}
K^{AB}= \left(
\begin{array}{cc}
0 & 0\\ 0 & k_1 p^{a}p^{b}
\end{array}
\right)\,,
\end{equation}
where
$k_1\ge0$
is a constant.

To determine $u^A$ we again look individually at the components in
spacetime and momentum space.
As before,
the spacetime component
$u^{\mu} = v^{\mu}$ by (\ref{e:uA}),
and $v^{\mu} = p^\mu$ by
(\ref{e:vdef}).
In contrast to the massive case, there can be nonzero
components of $u^A$ in the momentum space directions because
the momentum itself is an invariant vector.
The momentum direction components are thus given by
$u^{a}=k_2p^a$, where $k_2$ is a constant. Working in polar
coordinates the ``position'' vector $p^a$ on the cone $\mathbb{H}_0^3$
is simply $(p,0,0)$ where $p^2={p_0}^2$. Thus $u^A =
(p^0,p^1,p^2,p^3,k_2p,0,0)$ on $\mathbb{M}^4\times\mathbb{H}_0^3$.

Substituting the forms for $K^{AB}$ and $u^A$ into (\ref{e:diffequA})
we obtain the massless particle affine time equation:
\begin{eqnarray}
\frac{\partial\rho_{\lambda}}{\partial\lambda} &=&
\partial_A\left(K^{AB}n\partial_B\left(\frac{\rho_{\lambda}}{n}\right)-u^A\rho_{\lambda}\right)\nonumber\\
&=& -p^{\mu}\frac{\partial\rho_{\lambda}}{\partial x^{\mu}} +
k_1\frac{\partial}{\partial E}\left(E^3\frac{\partial}{\partial
E}\left(\frac{\rho_{\lambda}}{E}\right)\right)\nonumber\\
&& -
k_2\frac{\partial}{\partial E}\left(E\rho_{\lambda}\right)\,,\label{e:affine}
\end{eqnarray}
where we have replaced $p$ by energy $E=p$.

We see that the Lorentz invariance means that any diffusion in
photon momentum cannot change the direction of the photon and so
it always propagates on the light cone, at the speed of light.
However the energy of the photon does undergo a diffusion.
Notice also that there are two parameters, making this
a less powerful phenomenological model than the massive particle
model which has a single parameter. There is not only a diffusion
term but an independent drift term, arising from the
existence of an invariant vector on $\Lob_0$,
and we will see that this
leads to the existence of power law equilibrium solutions.
Note that taking the $m\rightarrow 0$ limit of (\ref{e:mtau})
would have resulted in (\ref{e:affine}) with $k_2=0$
because there is no invariant vector in the massive case.
(As is familiar in another context, the case of zero photon mass is thus,
here also, a sort of singular limit of the massive case.)

\subsection{Cosmic time process}

Again, in order to make contact with observations, we need to
obtain the cosmic time diffusion equation for massless particles,
for which we use the same  argument as in the massive case. First
we assume that $\lambda = 0$ at $t=0$. Let
\begin{equation}
J^{\lambda}(t,x^i,p^a,\lambda) = \rho_{\lambda}\;.
\end{equation}
We can then express the $t$ component of the current $J$ in terms of
$J^{\lambda}$, and the remaining components of the current in terms of
$J^t$.

\begin{eqnarray}
J^t(t,x^i,p^a,\lambda) &=& p J^{\lambda}\;;\\
J^{i}\left(t,x^i,p^a,\lambda\right) &=& \frac{p^i}{p}J^t\;;\\
J^{a}\left(t,x^i,p^a,\lambda\right) &=& -k_1 p^a p^b
{p\sin\theta}\,\partial_b\left(\frac{
J^t}{p^2\sin\theta}\right)\nonumber\\&& + \frac{J^t}{p}k_2 p^a\;.
\end{eqnarray}

\noindent
$J^a$ is proportional to $p^a$ and
in polar coordinates the vector $p^a = (p, 0, 0)$,  so
there is only one nonzero component of $J^{p^a}$
in the radial (energy) direction:
\begin{eqnarray}
J^{p}\left(t,x^i,p^a,\lambda\right) &=& -k_1 p\,\frac{\partial
J^t}{\partial p} + \left(2k_1 + k_2\right)J^t\,.
\end{eqnarray}
\noindent
The affine time of flight is unobservable so we integrate over it. Defining
\begin{eqnarray}
\bar{J}^t(t,x^i,p^a) &=& \int_0^\infty{J^t(t,x^i,p^a,\lambda)d\lambda},
\end{eqnarray}
\noindent
we integrate the other current components over $\lambda$ to obtain

\begin{eqnarray}
\bar{J}^{i}(t,x^i,p^a) &=& \frac{p^i}{p}\bar{J}^t\\
\bar{J}^{p}(t,x^i,p^a) &=& -k_1 p\frac{\partial \bar{J}^t}{\partial p}
+ \left(2k_1 + k_2\right)\bar{J}^t.
\end{eqnarray}
\noindent
Imposing the continuity equation, gives us the massless particle
cosmic time diffusion equation in terms of the scalar density
$\bar{J}^t$, which we rename $\rho_{t}$:
\begin{eqnarray}
\frac{\partial\rho_{t}}{\partial t} &=& -\partial_i J^i - \partial_a
J^a\nonumber\\  \label{e:diff2}
&=& -\frac{p^i}{E}\partial_i\rho_{t} - \left(k_1 +
k_2\right)\frac{\partial \rho_{t}}{\partial E} + k_1 E
\frac{\partial^2 \rho_{t}}{\partial E^2}\;,
\end{eqnarray}
where $E=p$ is the energy.

So, for a massless particle in a geometric optics approximation,
we expect that an underlying
discreteness can induce fluctuations in
the energy of the particle,
but without affecting the direction of propagation.
The diffusion governed by $k_1$ causes a
distribution of energies that is initially sharply peaked to
spread over time. The second constant $k_2$ results in an independent  drift of the
spectrum to higher or lower energies depending on its sign.

It is interesting
that negative values of $k_2$ allow for power law
equilibrium solutions of (\ref{e:diff2}). Set $\partial_\mu \rho_t =0$.
Then the equilibrium distributions satisfy
\be\label{e:equil}
- \left(k_1 +
k_2\right)\frac{\partial \rho_{t}}{\partial E} + k_1 E
\frac{\partial^2 \rho_{t}}{\partial E^2} = 0\;.
\ee
This has a power law solution
\be \rho_t \propto E^{\frac{2 k_1 + k_2}{k_1}}\;.
\ee
When the parameters are such that the exponent is
less than $-2$ (and so $k_2$ must be negative because $k_1$ is
positive)
 then this solution is normalisable if it is
cut off at small energies. We conjecture that if
$(2k_1 + k_2)/k_1 <-2$  any normalised
distribution will tend at late times to this
power law equilibrium solution at large energies.
This is interesting because physical processes that
result in power law distributions
across a wide energy range are few and far between
 --- the Fermi mechanism
of statistical acceleration of charged particles by random magnetic fields, proposed as
the source of high energy cosmic rays, is the only well-known mechanism.

Placing bounds on the parameters $k_1$ and $k_2$ is the next step.

\section{Bounding the constants $k_1$ and $k_2$}
\label{s:bounds}

In developing a phenomenological model, one aims to provide a model for
currently unexplained observations or suggest new observations that
might be made to test a theory.
But before proposing new observations, one should of course constrain
one's model as tightly as possible, based on what is already known.
Our model has two parameters: a positive diffusion constant $k_1$ and
a ``drift'' constant $k_2$, which may be either positive or negative.
To place the strongest bounds on the values of these parameters, it
seems sensible to look at photons that have been travelling for a very
long time and thus have had the maximum possibility to experience any
underlying discreteness.
The cosmic microwave background (CMB) seems an ideal testing ground in
this sense.
Not only are
its photons the
``oldest'' we can observe, but
its spectrum has been determined
with great precision.
Most of
the photons in the CMB have been ``free streaming''
for approximately $13.7$ billion years, or on the order of $10^{60}$
Planck times. When the universe became transparent
at recombination, they would have had a blackbody spectrum with a
temperature $3000\,K$ (see for example~\cite{Kolb:1990}).  Current
observations yield a temperature of $2.728\pm0.004 \,K$
and measure the spectrum to be Planckian (blackbody)
over the $2-21cm^{-1}$ frequency range
to within a
weighted rms deviation
of
only 50 parts per million (ppm) of
the peak brightness~\cite{Fixsen:1996nj}.
Since our diffusion would have
distorted the energy distribution,
the fact
that the CMB photons have travelled so far but remained so
perfectly thermal will allow us to constrain our parameters
very tightly.

\subsection{Simulations}

Our derivation of the massless cosmic time diffusion equation assumed
spacetime to be Minkowskian.  Therefore we will first consider a simplified
model that ignores the expansion of the universe, and consequently
assumes that, in the absence of diffusion, the temperature of the CMB
would remain constant from the surface of last scattering to today.
This will give us an order of magnitude bound on the parameters.  In
Section~\ref{s:expuni} the cosmic expansion will be incorporated.

The initial Planckian spectrum, expressed as a number density of
photons per unit spatial volume per unit energy, is
\begin{equation}
  \rho(E, t=0) = 8\pi\frac{E^2}{\exp\left(\frac{E}{T}\right) - 1}\,,
\end{equation}
with a temperature $T=2.728K$.  According to our model, this
distribution evolves via the homogeneous
massless cosmic time diffusion equation
\begin{equation} \label{e:homcos}
  \frac{\partial\rho_t}{\partial t} = -
  \left(k_1+k_2\right)\frac{\partial\rho_{t}}{\partial E} +
  k_1E\frac{\partial^2\rho_{t}}{\partial E^2}  \ .
\end{equation}
Using the \texttt{MATLAB} numerical pde solver \texttt{pdepe}, this
equation was integrated over a time interval equal to that since the
surface of last scattering.

Although only the $2$$-$$21cm^{-1}$ region of the spectrum is
needed to compare with the reported rms deviation, these evolutions
were run over a larger range of frequencies to capture more of the
spectrum and allow the implementation of a boundary condition at $E=0$.
What boundary condition is appropriate?
What happens to a photon as its momentum approaches zero?
Do photons leak away through the tip of the null cone in momentum space?
Physically, the photon concept employed by our model breaks down as the
wavelength tends to infinity, because the geometrical optics
approximation fails.
(Moreover, our affine parameter $\lambda$ fails to be well defined
physically, since it reaches infinity in a finite time if $E\to0$, and
thus cannot remain approximately constant over the photon wave-packet.)
This suggests that the so called ``absorbing boundary condition'',
$\rho(E)=0$, is appropriate at $E=0$, and this is what was used
in all our simulations.
In fact, the current is
\be
  J = (2k_1+k_2)\rho_t - k_1 E \partial\rho_t/\partial E \;,
\ee
so
(as long as $\partial\rho_t/\partial E$ remains finite)
any linear combination of $\rho_t=0$ with the ``reflecting boundary
condition'', $J=0$, is equivalent at $E=0$.

The evolved spectrum was converted from a number density per unit
volume per unit frequency to
a spectral radiance ---
energy per unit area per unit time per unit frequency per steradian
---  as used in the analysis of the COBE FIRAS data. This allows us to
compare the deviation from Planckian with the quoted 50ppm of the peak brightness.

A Planck spectrum was fit to the evolved spectral radiance
using the least squares method. By looking for the best fit Planck
spectrum rather than comparing with the initial $2.728K$ spectrum, we allowed for
the possibility that the diffusion changes the temperature of the CMB
in a way that may be reconciled with observation.
As it happens, we found that the temperature of the
best fit Planck spectrum was very close to the initial
temperature in cases where the deviation is within the allowed
tolerance. For example the choice of parameters $k_1 = 5\times 10^{-97}$
and $k_2=1\times 10^{-96}$ gives a best fit temperature of $2.7281 K$,
indistinguishable from the current observed temperature of $2.728\pm0.004 K$.
Finally the rms deviation between the fitted Planckian spectrum
and the evolved
spectrum in the
$2$$-$$21cm^{-1}$ frequency range (energy range
$4\times10^{-23}-4\times10^{-22}J$)
was
calculated with all points
weighted equally.
This result was compared to the allowed tolerance of
50 parts per million of the peak brightness.  This process
was repeated
for a range of values of the parameters $k_1$ and $k_2$.

\begin{figure*}[th]
\begin{center}
\subfigure[]{\label{f:k1only}
\includegraphics[width=0.4\textwidth]{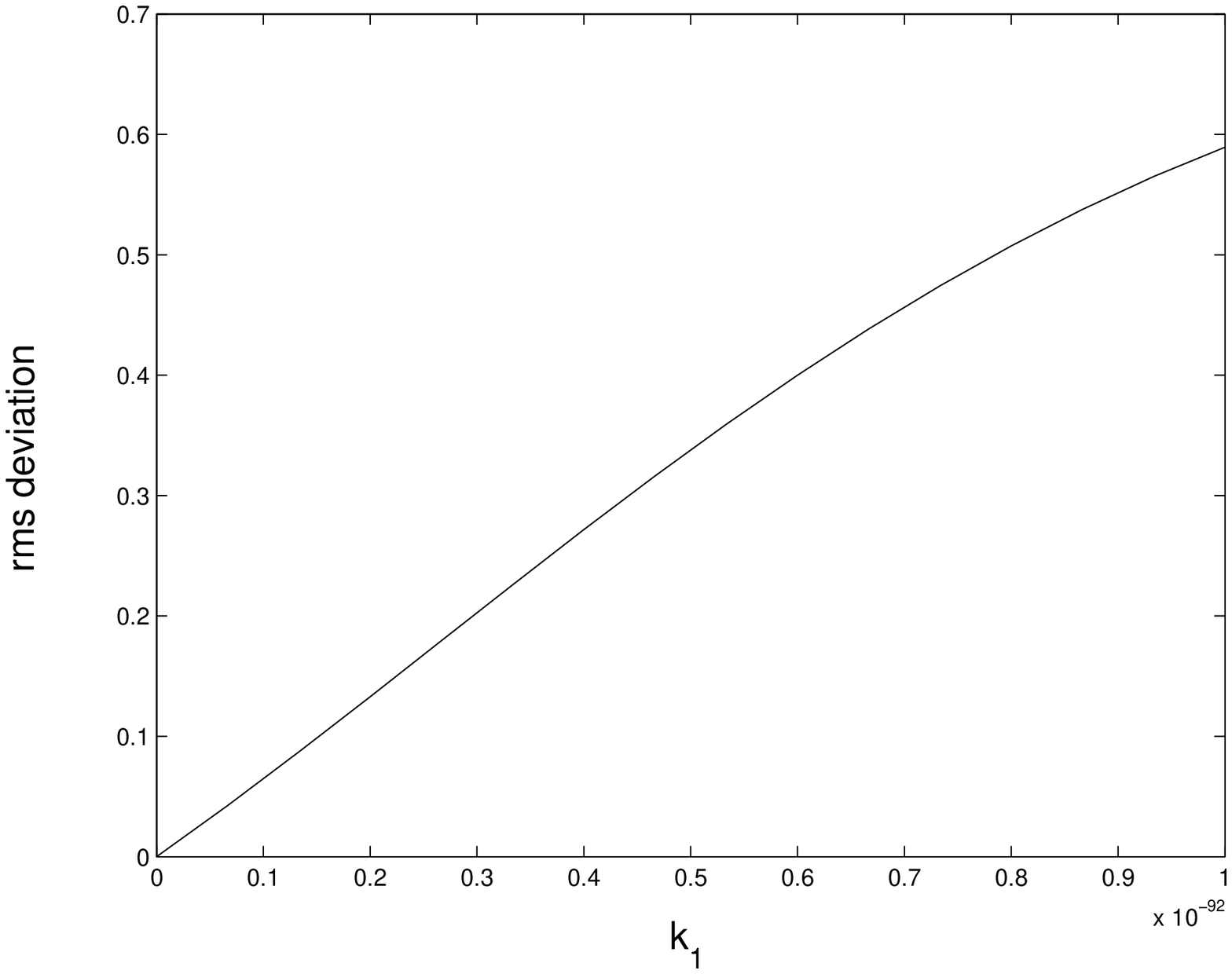}}
\hspace{8mm}
\subfigure[]{\label{f:k2only}
\includegraphics[width=0.41\textwidth]{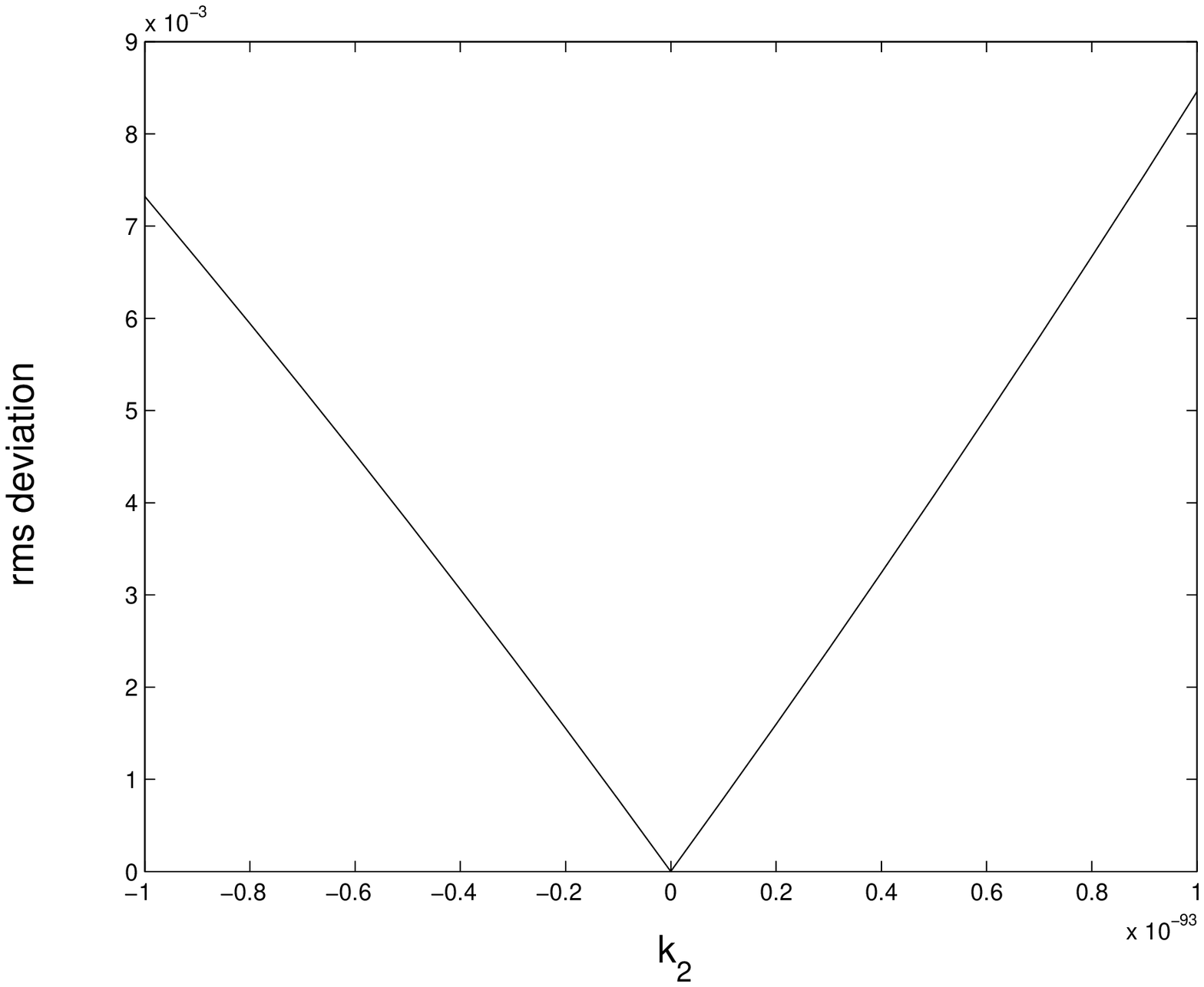}}
\caption{The rms deviation between the evolved spectrum and a bestfit
Planck spectrum (as a proportion of the spectrum peak):(a) varying
$k_1$ with $k_2 = 0$, and (b) varying $k_2$ with $k_1=0$.}
\end{center}
\end{figure*}

\begin{figure}[t]
\begin{center}
\includegraphics[width=0.35\textwidth]{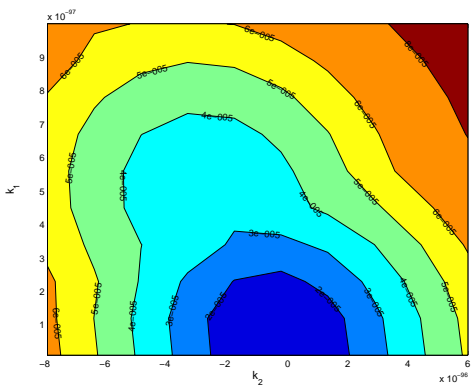}
\caption{The rms deviation of the simulated spectrum from
Planckian as a proportion of the spectrum peak, varying both
$k_1$ and $k_2$. Values of $k_1$ and $k_2$ within the
$5e$$-$$05$ contour give a spectrum that is Planckian to within 50ppm of the
peak. }
\label{f:k1k22pt7K}
\end{center}
\end{figure}

\subsection{Results}

We first place bounds on the diffusion and drift constants separately,
varying $k_1$ with $k_2 = 0$ and varying $k_2$ with $k_1 = 0$.

When $k_1=0$ we can solve the equation exactly:
\be
  \rho_t(E, t) = \rho_0(E - k_2 t)\,,
\ee
so the spectrum just translates at a constant speed.
For $k_2$ negative, this is inconsistent with the boundary
condition $\rho=0$ at $E=0$.  However, in this case one
can implement an absorbing boundary condition trivially:
simply cut off the translated distribution at $E=0$.
This is what we did to generate the solution plotted in Figure
\ref{f:k2only}.

One might be concerned that deviations within the allowed tolerance
would
be so small as to approach the level of the numerical errors
in the simulations. The exact solution for $k_1=0$ provides us with a
means of demonstrating that this is not the case~\footnote%
{We can also note that solving our diffusion equation in
 \texttt{Mathematica} using \texttt{NDSolve} yields the same results as
 discussed here, suggesting the bounds we obtain are robust,
 do not depend on the particular method of solving the equation,
 and are not a consequence of
 numerical error or a particular choice of integration step size.}.
When we compare the exact solution with the numerical solution for $k_1=0$
the errors introduced by the numerical integration can be seen to be
several orders of magnitude smaller than the deviation from Planckian.
For example if $k_2 = 4\times 10^{-96}$ the rms deviation from the best
fit Planck spectrum is $5\times 10^{-101}$ ($5\times 10^{-5}$ peak brightness)
for both the exact and the numerical solution. The rms deviation
\textit{between} the exact and numerical solution is $4\times 10^{-104}$.
If $k_2 = -4\times 10^{-96}$ the rms deviation from the best fit Planck spectrum
is also $5\times 10^{-101}$ ($5\times 10^{-5}$ peak brightness) while
the deviation between the exact and numerical solutions is again
$4\times 10^{-104}$.
This also demonstrates that the $\rho=0$ boundary condition we
imposed
on the numerical solution, although inconsistent with the
exact solution when $k_2<0$,
does not introduce noticeable
errors for the values
of $k_1$
that we are concerned with.

When $k_2=0$ with $k_1>0$ we can only solve the equation numerically.
The results for both cases are displayed in
Figures \ref{f:k1only} and \ref{f:k2only}.
We see that the deviation from
Planckian increases approximately linearly with increasing magnitude
of the parameters.
(Notice that figure \ref{f:k2only} was drawn from the exact solution, the
 graph taken from the numerical solution is indistinguishable.)
The
simulations suggest that for the deviation
from Planckian of the CMB to be within the allowed $5\times 10^{-5}$
of the peak brightness the diffusion constant $k_1$ must be less than
approximately $7\times10^{-97}$ if $k_2 = 0$, and the drift parameter
$k_2$ must fall within the range
$-4\times10^{-96}<k_2<4\times10^{-96}$ if $k_1 = 0$. Converting to SI
units we have the bounds:
\begin{eqnarray}
  &k_1&< 3\times10^{-44}kgm^2s^{-3}\,,\\
  -1\times10^{-43}<&k_2&<1\times10^{-43}kgm^2s^{-3} .
\end{eqnarray}

Similar bounds apply when we let both $k_1$ and $k_2$ be nonzero.
The general situation is displayed in Figure~\ref{f:k1k22pt7K},
from which one can read off
the values of $k_1$ and $k_2$ for which the deviation from blackbody
is less than $5\times10^{-5}$ of the peak brightness when we allow
both constants to vary.

In the units used here, the bounds on the parameters are very small.
However we can get a handle on where these numbers come from by
rescaling the energy,
setting $E' = s E$ with $s$ chosen so that $sT = 1$ when $T$ is the
CMB temperature. This means that $s\sim10^{32}$ in Planck units.
We rescale $\rho'=\rho/s$ so the initial spectrum is:
\be
  {\rho'}_0(E') = 8 \pi \frac{1}{s} \frac{{E'}^2}{ s^2 (e^{E'/T'} -1) }\,,
\ee
where $T' = sT = 1$.
If $k_2 =0$, then we can also
rescale the time, setting $t' = s \,k_1 t$ to obtain the diffusion
equation
\be
  \frac{\partial\rho'}{\partial t'} = - \frac{\partial\rho'}{\partial E'} +
  E'\frac{\partial^2\rho'}{\partial {E'}^2}\,.
\ee
If we now evolve $\rho'$ until it differs from ${\rho'}_0$ by 50 ppm and
take the value, $t'_f$ of $t'$ when this happens, $t'_f$ must,
for consistency with the data,
be greater than or
equal to $s \, k_1 t$ where $t$ is the age of the universe, and
so in Planck units $k_1 \le 10^{-60} 10^{-32} t'_f$.
We see that the order of magnitude bound found above will result if
$t'_f \sim 10^{-4}$,
which is indeed about the (rescaled) time at which one would have
expected the deviation to reach 50 ppm.
%
A similar order of magnitude estimate follows from the geometric
interpretation of our affine parameter $\lambda$ as an area, if one
notes that the product of the photon wavelength ($\sim1cm$) with the
Hubble radius is around $10^{32}10^{60}\sim 10^{92}$ in Planck units.

\section{Expanding universe}
\label{s:expuni}

In Section~\ref{s:bounds} we ignored the effect of the expansion of the
cosmos on the CMB and assumed that it
remained at a temperature of $\sim2.7K$ from the
surface of last scattering to today. This is of course not the
case. At the surface of last scattering the CMB had a temperature of
about $3000K$. As the universe expanded the individual photons were
stretched along with the space, and
correspondingly
 diluted, leaving us with the
$2.7K$ spectrum observed today. We will now show that the expansion
has essentially no effect on our model in the sense that the
distribution in the
expanding universe can be deduced easily from the nonexpanding one and
that the bounds derived from the nonexpanding simulation
change only slightly.

The redshifting effect of the expansion (but not the dilution) can be
added to the model by adding to $v$ a vector
which has a single component in the $E$ direction:
\be
  \Delta{v}^E = \frac{d E}{dt} = - E \frac{\dot{a}}{a}\;,
\ee
where $a(t)$ is the
cosmic scale factor.
This changes the continuity equation (\ref{e:continuity}) to
\begin{eqnarray}
\frac{\partial\rho_{t}}{\partial t} &=& -\partial_i J^i - \partial_a
J^a \\ &=& -\frac{p^i}{E}\partial_i\rho_{t} - \left(k_1 +
k_2\right)\frac{\partial \rho_{t}}{\partial E}\nonumber\\&& + k_1 E
\frac{\partial^2 \rho_{t}}{\partial E^2} + \frac{\dot{a}}{a} \frac{\partial}
{\partial E}(\rho_t E)\;. \label{e:withexpan}
\end{eqnarray}

A solution of this equation, for $k_1=k_2 =0$ is
\begin{equation}\label{rhozero}
{\rho}_0\left(E,t\right) = 8\pi \frac{a^3}{a_0^3} \frac{E^2}
{e^{\frac{E}{T_0}\frac{a}{a_0}} -1} \;,
\end{equation}
where $a_0$ is the scale factor at time $t_0$. If we multiply this
distribution by $\frac{a_0^3}{a^3}$, which dilutes the photons
according to the expansion, it becomes exactly the Planck
distribution for temperature
$T = T_0 \frac{a_0}{a}$.

If we define a new variable $\widetilde{E} = \frac{a}{a_0}E$ and
a new density function
$\widetilde{\rho}(\widetilde{E}) =  \frac{a_0}{a} \rho(E)$
(this
being just the
transformation of a scalar density under a rescaling of
coordinates: $\rho\, dE = \widetilde\rho\, d\widetilde E$)
 the distribution $\rho_0(E,t)$ (\ref{rhozero})
becomes
\begin{eqnarray}
\widetilde{\rho}_0\left(\widetilde{E},t\right)
 = 8 \pi
\frac{\widetilde{E}^2}{\exp\left(\frac{\widetilde{E}}{T_0}\right) - 1}\,,
\end{eqnarray}
which is constant in time.

We now transform our diffusion equation to the rescaled quantities
$\widetilde{\rho}$ and $\widetilde{E}$.

Starting with (\ref{e:withexpan}), we have:
\begin{align}
LHS &=
\left( \frac{\partial}{\partial t} + \frac{\dot{a} \widetilde{E}}{a}
\frac{\partial}{\partial \wE}\right)
\left(a \wrho\right)\\
 & = \dot{a}\wrho + a \dot{\wrho} + \dot{a} \wE \wrho'\\
RHS & = - (k_1 + k_2) a\left(a \wrho\right)'
+ k_1 a^2 \wE \wrho'' + \dot{a} \left( \wrho \wE\right)' \\
&= -(k_1+k_2) a^2 \wrho' + k_1 a^2 \wE \wrho'' + \dot{a} \wrho +
\dot{a}\wE \wrho'\,,
\end{align}
where dot denotes time derivative and prime denotes derivative
with respect to
$\wE$.
This gives
\begin{equation}
  \frac{\partial\widetilde{\rho}}{\partial t}
  = -(k_1+k_2)a\frac{\partial\widetilde{\rho}}{\partial\widetilde{E}}
  + k_1 a
  \widetilde{E}\frac{\partial^2\widetilde{\rho}}{\partial\widetilde{E}^2}.
\end{equation}

Choosing $t'$ such that $\frac{dt'}{dt} = a$, we obtain
\be
\frac{\partial \widetilde \rho}{\partial t'} =
-(k_1+k_2) \frac{\partial}{\partial \widetilde E} \widetilde{\rho}
+ k_1 \widetilde{E} \frac{\partial^2}{\partial {\widetilde{E}}^2} \widetilde{\rho}\,,
\ee
which is the same as
(\ref{e:homcos}), the nonexpanding diffusion equation.

That we can find expanding solutions from static
ones is due to the scale-invariance of the
null cone $\Lob_0$~:
its geometrical structures are invariant under
$E\rightarrow \widetilde{E} = \textrm{const} \times E$.

For a matter dominated FRW universe $a\sim t^{2/3}$ i.e. $a(t) =
t^{2/3}/t_0^{2/3}$, where $t_0$ is the current value of $t$ (and
the current value of $a$ is 1).
We have $\frac{dt'}{dt} = a$ which integrates to
\be
t' = \frac{3}{5}\frac{t^{\frac{5}{3}}}{t_0^{\frac{2}{3}}} +
\textrm{const}\;.
\ee
If the range for $t$ is $10^{60}$ then the range for
$t'$ is 3/5 of this.
So the simulations we would need to do for the
expanding case are the same as for the nonexpanding case but
for only 3/5 of the time. This doesn't  affect the order of
magnitude of the bounds.

\section{Discussion}

The work presented here illustrates the familiar fact that
considerations of symmetry can bring forth a fairly unique
phenomenological model, even when relatively little is known about the
deeper reality the model is meant to represent.  Starting from the
assumption of an underlying spatiotemporal discreteness that
nevertheless respects Lorentz invariance in the continuum approximation,
we argued that particle momenta would be subject to stochastic
variations, and that if these variations were small, their effects would
be describable on large scales as a diffusion in momentum space.  The
assumption of Lorentz symmetry lends the resulting models their power
(by limiting the number of parameters), and it sets them apart from the
majority of quantum gravity phenomenological models, which break Lorentz
invariance.

For particles without internal degrees of freedom, we have seen that
even in the absence of a definite microscopic theory, an effective
diffusion model can be derived based on the assumed invariance alone.
One can also imagine applying this idea more generally, including for
example the polarisation of photons or neutrinos.

In the case of massive particles, if one of the explicit microscopic
models is fixed upon, then the diffusion strength, $k$, will be
a function of the forgetting time (number).  This forgetting time
sets the
scale shorter than which the dynamics is nonlocal: at much
larger scales the model is effectively local.
 In more realistic, more quantal models,
the diffusion scale
might also depend on such dimensionless numbers as the ratio
of the mass of the particle to the Planck mass and
properties of the particle's wave packet.
The same possibilities exist for the massless case.
Thus we would
expect the diffusion and drift parameters, $k_1$ and $k_2$,
to depend on some non-locality scale in the underlying physics,
and they could also depend on features of the wave packet
associated with the photon, for example the ratio of the (peak)
wavelength to the length or the packet.
In seeking an underlying
model of photons, the Lorentz invariant, nonlocal
D'Alembertian that has recently been
discovered for scalar field propagation on
causal set backgrounds \cite{Henson:2006kf,Sorkin:2007qi} could
be valuable.
Using it to evolve a wave packet of a massless scalar field, one
could ask whether the resulting propagation exhibited any momentum
diffusion or drift, and if so, what sets the scale of these phenomena.

The parameters of our model are constrained by the blackbody character
of the CMB radiation.  Since most observational astrophysics and
cosmology relies on electromagnetic radiation, there are a host of other
observations that could also be brought to bear, given that our model
entails a broadening of spectral lines as well as a distance-dependent
shift in energy.  For example, if the diffusion constant were set to
zero, it would be easy to work out how the drift would affect absorption
spectra from distant objects.  It seems likely however that the bounds
set here will be among the most stringent.

The models discussed here describe free point particles. Although we expect that composite objects would be less affected by the underlying discreteness (for example, a helium ion would swerve less than a proton) we can not make conclusive statements without a causal set model for interacting particles.

\begin{acknowledgments}
We thank Joe Henson for invaluable help with the expanding case
and Carlo Contaldi for useful discussions.
LP is supported by the Tertiary Education Commission of NZ (TAD1939).
FD is supported in part by Marie Curie Research and Training Network
``Random Geometry and Random Matrices:
From Quantum Gravity to Econophysics''
(MRTN-CT-2004-005616)
and the Royal Society grant IJP - 2006/R2.
Research at Perimeter Institute for Theoretical Physics is
supported in part by the Government of Canada through NSERC and
by the Province of Ontario through MRI.  This research was
partly supported by NSF grant PHY-0404646.

\end{acknowledgments}

\bibliography{refs}
\end{document}